\documentclass{aa}
\usepackage{graphicx}
\begin{document}

   \title{Chemical abundances in spiral and irregular galaxies}

   \subtitle{O and N abundances derived from global emission--line spectra}

\author{Leonid S.~Pilyugin \inst{1,2},
        Thierry Contini \inst{2},
        Jos\'{e} M. V\'{\i}lchez \inst{3}}

  \offprints{L.S.~Pilyugin}

   \institute{   Main Astronomical Observatory
                 of National Academy of Sciences of Ukraine,
                 27 Zabolotnogo str., 03680 Kiev, Ukraine
                 (pilyugin@mao.kiev.ua)
\and
                 Laboratoire d'Astrophysique de l'Observatoire Midi--Pyr\'{e}n\'{e}es -- UMR 5572,
                 14 avenue E. Belin, F-31400 Toulouse, France
                 (contini@ast.obs-mip.fr)
                 \and
                 Instituto de Astrof\'{\i}sica de Andaluc\'{\i}a,
                 CSIC, Apdo, 3004, 18080 Granada, Spain
                 (jvm@iaa.es)
                 }

\date{Received 28 November 2003 / accepted 04  April 2004}

\abstract{
The validity of oxygen and nitrogen abundances derived from the global
emission-line spectra of galaxies via the P--method has been investigated
using a collection of published spectra of individual H\,{\sc ii} regions in
irregular and spiral galaxies.  The conclusions of Kobulnicky, Kennicutt \&
Pizagno (1999) that global emission-line spectra can reliably indicate the
chemical properties of galaxies has been confirmed. It has been shown
that the comparison of the global spectrum of a galaxy with a collection of spectra
of individual H\,{\sc ii} regions can be used to
distinguish high and low metallicity objects and to estimate accurate
chemical abundances in a galaxy. The oxygen and nitrogen abundances in samples of
UV-selected and normal nearby galaxies have been
determined. It has been found that the UV-selected galaxies
occupy the same area in the N/O -- O/H diagram as individual H\,{\sc ii} regions
in nearby galaxies. Finally, we show that intermediate-redshift galaxies
systematically deviate from the metallicity -- luminosity trend of
local galaxies.
}

\titlerunning{O and N abundances derived from global spectra }

\authorrunning{L.S.~Pilyugin et al.}

\maketitle

\section{Introduction}

\begin{figure}
\resizebox{1.00\hsize}{!}{\includegraphics[angle=000]{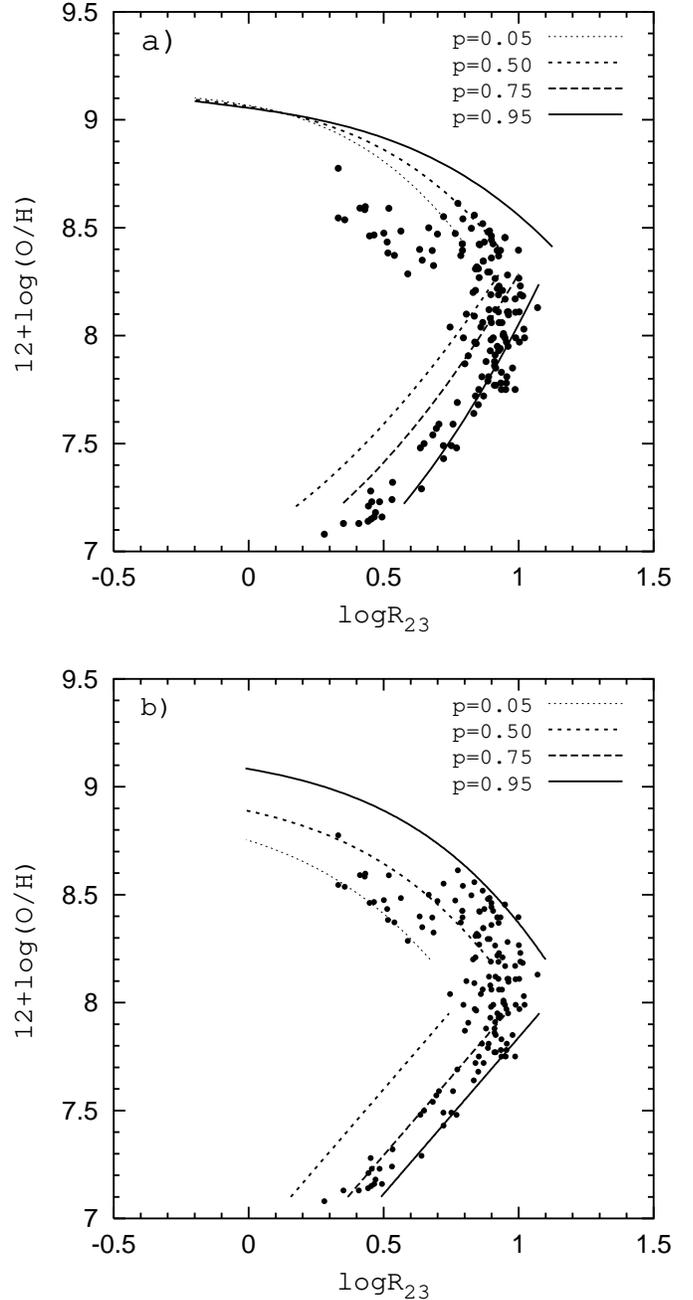}}
\caption{ The "strong line -- oxygen abundance" calibrations. 
{\bf a)} 
Points correspond to H\,{\sc ii} regions with oxygen abundances determined 
using direct measurements of $T_{\rm e}$ -- method (from the compilation of 
Pilyugin 2000, 2001a).
The lines show the predictions of the two-dimensional theoretical (model) 
calibration of Kobulnicky, Kennicutt \&
Pizagno (1999) parameterized for different values of the excitation 
parameter $P$.  Every line is labeled with
the corresponding value of the excitation parameter $P$.
{\bf b)} The lines are the predictions of the two-dimensional empirical 
calibrations obtained 
for the lower branch (12+log(O/H) $<$ 8.0; Pilyugin 2000, 2001c) and for the 
upper branch (12+log(O/H) $>$ 8.2; Pilyugin 2001a). Each O/H -- $R_{\rm 23}$ relation 
is labeled with the corresponding value of the excitation parameter $P$. 
The points are the same as in panel {\bf a)}.
}
\label{figure:r0}
\end{figure}

Investigating the variations of chemical properties among galaxies is very
important to understand their structure and evolution.
Abundance determinations in spiral galaxies are based on the
spectrophotometry of individual H\,{\sc ii} regions.
Good spectrophotometry of H\,{\sc ii} regions is now available for a large
number of galaxies, and the reliability of chemical abundance determinations 
is defined mainly by the method used for abundance determinations in H\,{\sc ii} regions.
Accurate abundances in H\,{\sc ii} regions can be derived through the classical $T_{\rm e}$
-- method if measurements of temperature-sensitive line ratios are available.
Unfortunately, in oxygen-rich H\,{\sc ii} regions the temperature-sensitive lines such
as [OIII]$\lambda$4363 are often too weak to be detected. For such H\,{\sc ii}
regions, the relation between strong oxygen line intensities and oxygen
abundances is used for the abundance determination. The early calibrations
were one-dimensional (Pagel et al. 1979; Edmunds \& Pagel 1984; McCall
et al. 1985; Dopita \& Evans 1986; Zaritsky et al. 1994), i.e. a relation of
the type $O/H = f(R_{\rm 23})$ was used, with 
$R_{\rm 23}$ = ($I_{[OII] \lambda 3727+ \lambda 3729}$ + 
 $I_{[OIII] \lambda 4959+ \lambda 5007}) /I_{H\beta }$.
It has been shown (Pilyugin 2000; 2001a,b)
that the oxygen abundances derived with the one-dimensional
calibrations are affected by a systematic error. The origin of this systematic error
is evident. In a general case, the intensity of oxygen emission lines in H\,{\sc ii} 
regions depends not only on the oxygen abundance but also on the
physical conditions (hardness of the ionizing radiation 
and geometrical factors). Thus, the physical conditions in H\,{\sc ii} regions 
(e.g. via the electronic temperature in the $T_{\rm e}$ -- method) should 
be taken into account to derive accurate oxygen abundances from emission line 
intensities. In one-dimensional calibrations the physical
conditions in H\,{\sc ii} regions are ignored. Starting from the idea of McGaugh (1991)
that the strong oxygen lines contain the information needed to determine
accurate abundances in H\,{\sc ii} regions, it has been shown
that the physical conditions in H\,{\sc ii} regions can be estimated and taken 
into account via the excitation parameter $P$ (see Fig.~\ref{figure:r0}). 
A two-dimensional or parametric calibration (the $P$ -- method) has been 
proposed (Pilyugin 2000; 2001a,c). A more general relation
of the type $O/H = f(P, R_{23})$ is used in the $P$ -- method, compared to the
relation of the type $O/H = f(R_{23})$ used in one-dimensional calibrations 
(where $R_{\rm 23}$ =$R_{\rm 2}$ + $R_{\rm 3}$,
$R_{\rm 2}$ = $I_{[OII] \lambda 3727+ \lambda 3729} /I_{H\beta }$,
$R_{\rm 3}$ = $I_{[OIII] \lambda 4959+ \lambda 5007} /I_{H\beta }$,
and P = $R_{\rm 3}$/$R_{\rm 23}$).
The oxygen abundances derived via the $P$ -- method are in agreement with
the abundances determined using the $T_{\rm e}$ -- method over the full range of
abundances (see Fig.~\ref{figure:r0}, and Melbourne et al. 2004). This allows us to conclude
that the $P$ -- method provides reliable abundance determinations in H\,{\sc ii} regions.

The correlations between oxygen abundance traced by the individual H\,{\sc ii}
regions and macroscopic properties of spiral galaxies have been investigated
by numerous authors (Garnett \& Shields 1987; Vila-Costas \& Edmunds 1992; Zaritsky,
Kennicutt \& Huchra 1994; Garnett 2002; among others). The early one-dimensional 
$R_{\rm 23}$ calibrations were used in these studies for the oxygen abundance determination.
Therefore, the results of these studies are not beyond question and
must be revised.
A compilation of published spectra of H\,{\sc ii} regions in spiral galaxies
has been carried out in our previous paper (Pilyugin, V\'{\i}lchez \& Contini
2004; hereafter Paper I). The oxygen and nitrogen abundances in all the
H\,{\sc ii} regions were determined in the same way, using the $P$--method.
The correlations between oxygen abundance and macroscopic properties of spiral
galaxies have been investigated.

In a pioneering study, Kennicutt (1992) published the global, i.e. spatialy
unresolved, spectrophotometry of a sample of 90 galaxies spanning the entire
Hubble sequence.
Kobulnicky, Kennicutt \& Pizagno (1999) have shown that the oxygen abundances
determined through the $R_{\rm 23}$--method of Pagel et al. (1979) using the
global emission-line spectra are in agreement with oxygen abundances measured
at 0.4 isophotal radius using individual H\,{\sc ii} regions.
They concluded that the global emission-line spectra can reliably indicate
the chemical properties of galaxies. Based on this conclusion, Melbourne \&
Salzer (2002) have estimated the oxygen abundance for a large sample of
519 star-forming emission-line galaxies from the KPNO International Spectroscopic
Survey (KISS) and derived the luminosity--metallicity ($L-Z$) relation.
During the last few years, chemical abundances have been
derived from the global spectra of a large sample of early-type (e.g.
Wegner et al. 2003) and star-forming galaxies (Melbourne \& Salzer 2002;
Contini et al. 2002; Tremonti et al. 2003; Lamareille et al. 2004).
Contini et al. (2002) have determined the chemical abundances in a sample
of UV-selected local and intermediate-redshift galaxies from their global spectra.
The N/O -- O/H diagram obtained by Contini et al. (2002) for these galaxies
is not in agreement with the one derived for individual  H\,{\sc ii} regions
in normal spiral and irregular galaxies (e.g. Pilyugin, Thuan \& V\'{\i}lchez
2003). For a given metallicity, a significant number of the UV-selected galaxies
show lower N/O abundance ratios as compared to the individual H\,{\sc ii} regions.

The N/O -- O/H diagram is a key relation for understanding the origin of
nitrogen and the star formation history of galaxies.
Therefore it is very important to understand the origin of the
low N/O ratios derived in a large fraction of UV-selected galaxies. It has been suggested
(Contini et al. 2002; Mouhcine \& Contini 2002) that the low N/O abundance
ratios in UV-selected galaxies are due to the fact that UV-selected galaxies
are observed at a special stage of their evolution. They have just undergone a
powerful starburst which enriched their interstellar medium in oxygen. In these
objects nitrogen may not have been completely released.

It is possible, however, that the low N/O ratios in UV-selected galaxies are 
caused by systematic errors in the determination of N/O abundance ratios.
Indeed, the abundances in UV-selected galaxies were determined from the global
spectra using the calibration of Kobulnicky, Kennicutt \& Pizagno (1999, see 
Fig.~\ref{figure:r0}a). This calibration is
based on the grid of H\,{\sc ii} region models of McGaugh (1991).
The N/O -- O/H diagram of Pilyugin, Thuan \& V\'{\i}lchez (2003) is based on
the abundances derived in individual H\,{\sc ii} regions using the $P$--method
(or via the classical $T_{\rm e}$--method when possible).
It should be stressed that "strong lines -- oxygen abundance" calibrations do not
form an uniform family. One should clearly recognize that there are two different
types of calibrations. The calibrations of the first type are empirical, i.e. based 
on oxygen abundances in H\,{\sc ii} regions determined via the $T_{\rm e}$--method. 
The calibrations of Pilyugin are two-dimensional calibrations of this type. 
The calibrations of the second type are ``theoretical'' (or model) calibrations, i.e. 
based on the grid of photoionization models of H\,{\sc ii} regions. The calibration 
of Kobulnicky, Kennicutt \& Pizagno (1999) is the two-dimensional calibration of this 
type.

Inspection of Fig.~\ref{figure:r0}
shows that the discrepancy between the calibration of Kobulnicky, Kennicutt 
\& Pizagno (1999) and the calibrations of Pilyugin is very small for the very
low-metallicity (12+log(O/H) $\sim$ 7.3), high-excitation H\,{\sc ii} regions. 
But the discrepancy increases with increasing metallicity and with decreasing
excitation parameter, reaching the value of $\Delta$log(O/H) $\sim$ 0.15 dex
for H\,{\sc ii} regions with 12+log(O/H) $\sim$ 7.9. The agreement between the
calibrations of Kobulnicky, Kennicutt \& Pizagno (1999) and those of Pilyugin 
disappears for high-metallicity H\,{\sc ii} regions that lie on the upper branch of the
O/H -- $R_{\rm 23}$ relation (see Pilyugin 2003 for a more detailled discussion).
The calibrations of Kobulnicky, Kennicutt \& Pizagno (1999) do not
reproduce the $T_{\rm e}$-based abundances for
high-metallicity H\,{\sc ii} regions that lie on the upper branch of the
O/H -- $R_{\rm 23}$ relation. Then,
it is possible that the low N/O ratios in UV-selected galaxies is
caused by systematic errors in the determination of abundances.
Therefore, to establish whether the low N/O abundance ratios of
UV-selected galaxies are real or artificial, it is necessary to derive abundances
for UV-selected galaxies through the $P$--method. This is one of the main goals of this work.

The plan of the paper is the following. In the next Section, the validity of
the abundances derived from the global spectra through the $P$--method is examined. 
The N/O -- O/H diagram for UV-selected galaxies is built and compared with the data 
from Pilyugin, Thuan \& V\'{\i}lchez (2003) in
Section 3. In Section 4 the luminosity--metallicity relation for the
UV-selected galaxies is derived and compared with samples of
normal nearby spiral galaxies (Jansen et al. 2000; Paper I) and with the
luminosity -- metallicity  relation obtained for intermediate-redshift galaxies.
Section 5 is a brief conclusion.

\section{On the validity of abundances derived from global spectra of galaxies}

\begin{figure}
\resizebox{1.00\hsize}{!}{\includegraphics[angle=270]{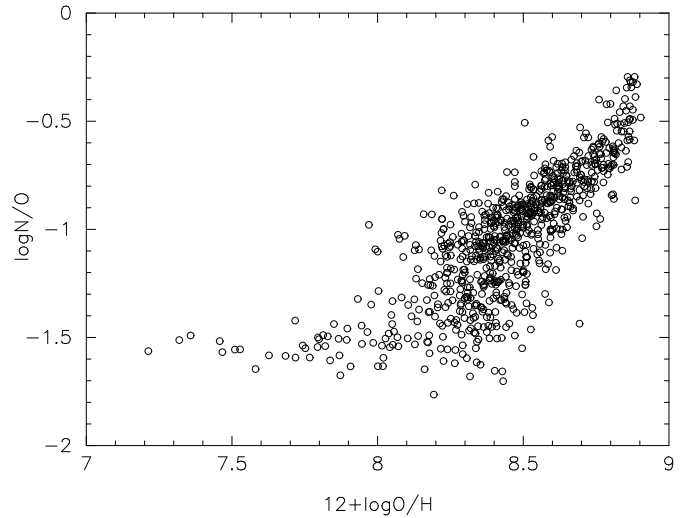}}
\caption{ The N/O vs. O/H diagram for our collection of H\,{\sc ii}
regions in spiral and irregular galaxies.
}
\label{figure:fxxx}
\end{figure}

\begin{figure}
\resizebox{1.00\hsize}{!}{\includegraphics[angle=270]{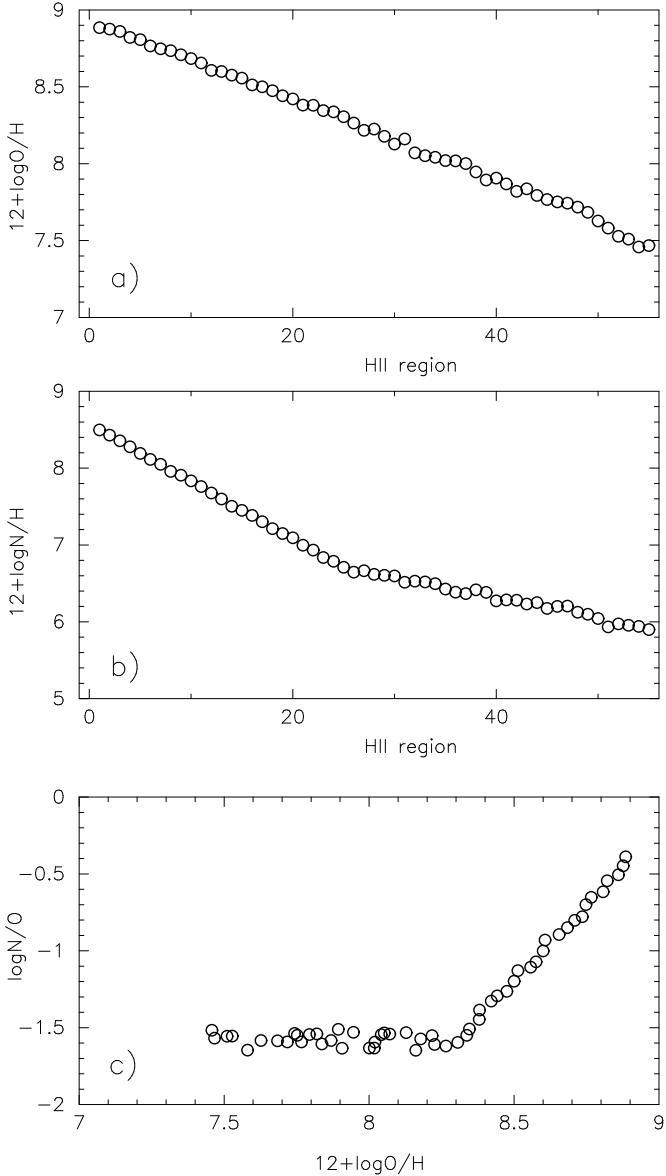}}
\caption{Oxygen (a) and nitrogen (b) abundances as well as the N/O -- O/H
diagram (c) for a basic subset of H\,{\sc ii} regions selected for the
construction of an artificial galaxy.
}
\label{figure:hiisel}
\end{figure}

\begin{figure}
\resizebox{1.00\hsize}{!}{\includegraphics[angle=270]{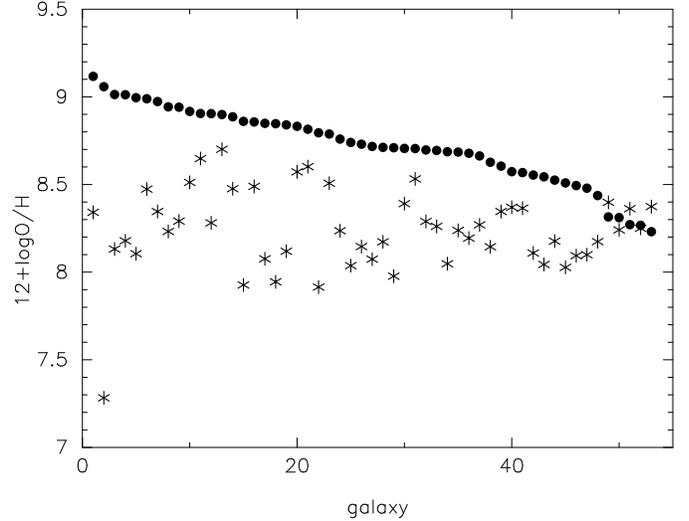}}
\caption{Central intersect oxygen abundances (filled circles) and oxygen
abundances at isophotal radius (asterisks) for the set of spiral galaxies
from Paper I. The galaxies are plotted in decreasing central O/H order.
}
\label{figure:GranadZ}
\end{figure}

\begin{figure}
\resizebox{1.00\hsize}{!}{\includegraphics[angle=270]{aa745f05.ps}}
\caption{The maximum ("central intersect") oxygen abundance (large open circles),
the minimum ("at the isophotal radius") oxygen abundance (open triangles), and
the oxygen abundance derived from the global spectra (filled circles) for
models of the basic serie.
}
\label{figure:agoh}
\end{figure}

\begin{figure}
\resizebox{1.00\hsize}{!}{\includegraphics[angle=000]{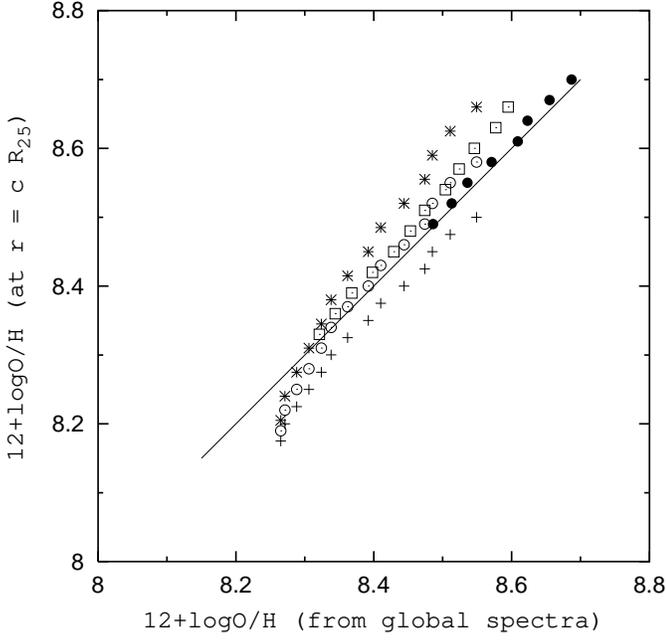}}
\caption{Oxygen abundances derived from the global spectra versus oxygen abundances
at $r=c\times R_{\rm 25}$.
The pluses correspond to abundances at $r=0.5\times R_{\rm 25}$ for basic
series of models with $Z_{\rm min}$ = 8.12.
The open circles corresponds to abundance at $r=0.4\times R_{\rm 25}$ for basic
series of models with $Z_{\rm min}$ = 8.12.
The asterisks correspond to abundances at $r=0.3\times R_{\rm 25}$ for basic
series of models with $Z_{\rm min}$ = 8.12.
The filled circles correspond to abundances at $r=0.4\times R_{\rm 25}$ for basic
series of models with $Z_{\rm min}$ = 8.42.
The open squares correspond to abundances at $r=0.4\times R_{\rm 25}$ for
series B of models with $Z_{\rm min}$ = 8.30.
}
\label{figure:z04zg}
\end{figure}

\begin{figure}
\resizebox{1.00\hsize}{!}{\includegraphics[angle=000]{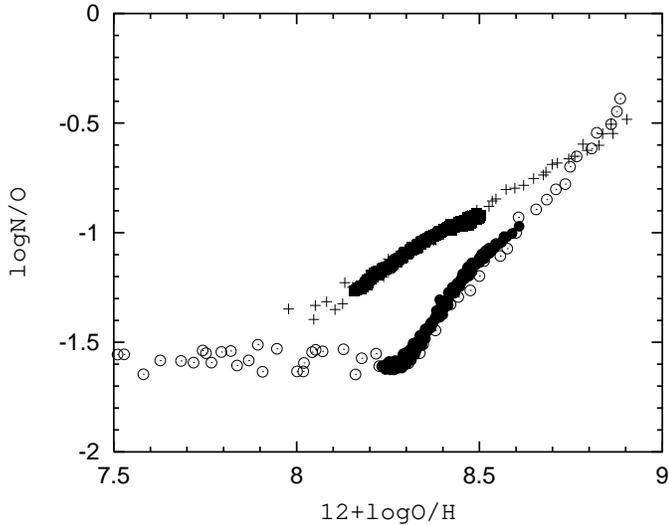}}
\caption{N/O vs. O/H diagram. The filled circles are the global abundances
derived for the models of the basic serie from Fig.~\ref{figure:agoh},
the open circles are the corresponding selected subsets of individual
H\,{\sc ii} regions.
The filled squares are the global abundances derived for the models of serie B
with $Z_{\rm min}$ = 8.1,
the pluses are the corresponding selected subsets of individual
H\,{\sc ii} regions.
}
\label{figure:agno}
\end{figure}

\begin{figure}
\resizebox{1.00\hsize}{!}{\includegraphics[angle=270]{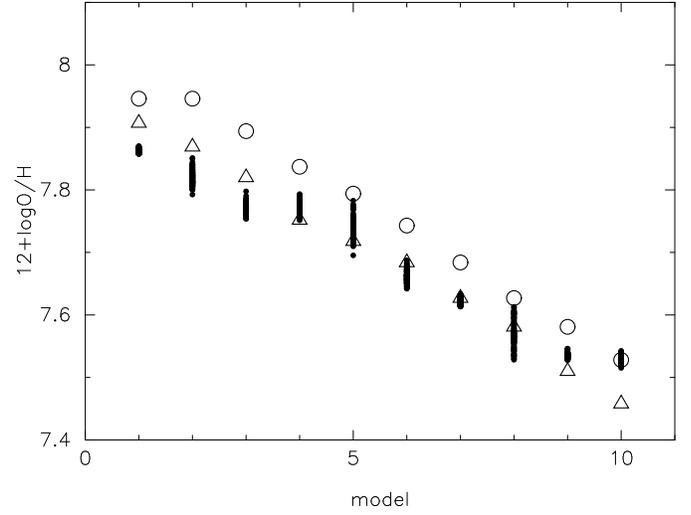}}
\caption{The maximum ("central intersect") oxygen abundance (large open circles),
the minimum ("at the isophotal radius") oxygen abundance (open triangles), and
the oxygen abundance derived from the global spectra (filled circles) for
models of low-metallicity galaxies.
}
\label{figure:agohlow}
\end{figure}

\begin{figure}
\resizebox{1.00\hsize}{!}{\includegraphics[angle=270]{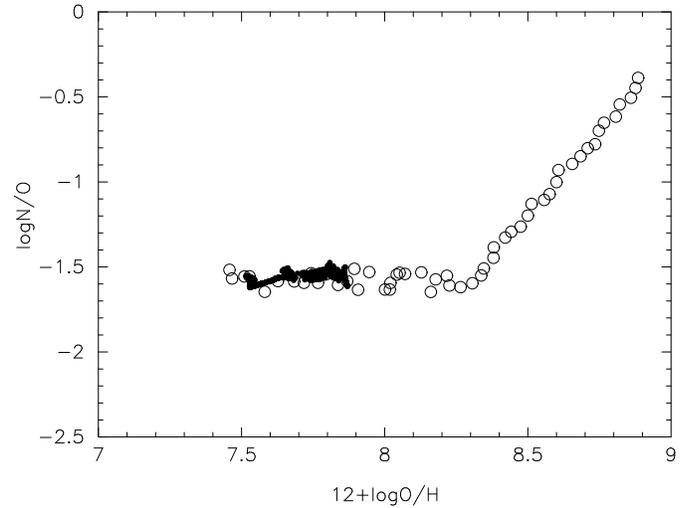}}
\caption{The N/O vs. O/H diagram. The filled circles are the global abundances
derived for the models of low-metallicity galaxies from Fig.~\ref{figure:agohlow}.
The open circles are the selected subsets of individual H\,{\sc ii} regions.
}
\label{figure:agnolow}
\end{figure}

The validity of abundances derived from the global spectrum of galaxies, 
using the $P$--method, will be examined in the following way. 
A model spectrum for a galactic disk is built using individual spectra 
of H\,{\sc ii} regions. The disk is divided into $n$ rings. The global spectrum 
of such an ``artificial'' galaxy is built by summing the spectra of $n$ rings. 
For the sake of clarity, the spectrum of an individual ring is represented by 
the spectrum of a single H\,{\sc ii} region. The compilation of
published line intensities for a large sample of individual  H\,{\sc ii} regions
in spiral and irregular galaxies (Paper I; Pilyugin, Thuan \& V\'{\i}lchez
2003) has been used to perform this work. The N/O versus O/H diagram for our 
collection of H\,{\sc ii} regions is shown in Fig.~\ref{figure:fxxx}. 
We select the basic subsample
of H\,{\sc ii} regions in such a way that this subsample reproduces a sequence
with a more or less monotonic decrease of oxygen and nitrogen abundances
(with a step $\Delta$log(O/H) $\sim$ 0.02dex) and reflects the general behaviour of the
nitrogen-to-oxygen ratio (Fig.~\ref{figure:hiisel}). This subsample of
H\,{\sc ii} regions has been used to build the basic sequence of model galaxy 
spectra.
The large scatter observed in the N/O vs. O/H diagram (Fig.~\ref{figure:fxxx}) 
is due to the fact that data for spiral galaxies of different morphological
type are presented. Early-type spirals (Sa, Sab) show high N/O ratios 
whereas late-type spirals (Sd, Sdm) have a lower N/O ratio at a given O/H 
(Pilyugin, Thuan \& V\'{\i}lchez 2003). The basic subsample of selected  
H\,{\sc ii} regions is close to the lower envelope in the N/O vs. O/H diagram. 
This subsample thus corresponds to spiral galaxies of late morphological types.
Galaxy model spectra with other subsamples of H\,{\sc ii} regions have been 
also constructed.

The oxygen abundance in the disk of spiral galaxies can be specified by the
maximum (or central intersect oxygen abundance), $Z_{\rm max}$ = 12 + $\log$(O/H)
(at $r=0$), and the minimum oxygen abundance (or oxygen abundance at the isophotal
radius), $Z_{\rm min}$ = 12 + $\log$(O/H) (at $r=R_{\rm 25})$.
The radial oxygen abundance distributions in the disk of 53 well-studied spiral
galaxies have been determined in Paper I using the abundances in
individual H\,{\sc ii} regions derived via the $P$--method.
The extrapolated central intersect
values of O/H and oxygen abundances at the isophotal radius $R_{\rm 25}$
for these galaxies are presented in Fig.~\ref{figure:GranadZ}. The galaxies
are plotted in decreasing central O/H order.
Fig.~\ref{figure:GranadZ} shows that both the maximum and the minimum oxygen
abundances can significantly vary from galaxy to galaxy.
Therefore we have built a basic sequence of galaxy model spectra with different 
central intersect oxygen abundances $Z_{\rm max}$ =
12 + $\log$(O/H) (at $r=0$) and a fixed oxygen abundance at the isophotal radius,
$Z_{\rm min}$ = 12 + $\log$(O/H) (at $r=R_{\rm 25}$) $\sim$ 8.1.
It should be emphazised that the goal of the present section is to test the
credibility of abundances derived from global spectra via the $P$--method
but not to reproduce all the galaxies from Paper I. The value of $Z_{\rm min}$
was choosen in a such way that
the basic sequence of model spectra reproduces a similar fraction of galaxies 
in which the bulk of H\,{\sc ii} regions belong to the upper branch of the 
O/H -- $R_{\rm 23}$ diagram (at high values of $\Delta$Z = $Z_{\rm max}$ -- $Z_{\rm min}$)
and galaxies in which a significant number of H\,{\sc ii} regions belong to the lower
branch of the O/H -- $R_{\rm 23}$ diagram (at low values of $\Delta$Z = 
$Z_{\rm max}$ -- $Z_{\rm min}$).
Other sequences of galaxy model spectra (with other values of
$Z_{\rm min}$, and/or with other subsamples of H\,{\sc ii} regions) have been also
constructed. It has been found that the results obtained for other sequences
of models do not differ from the ones obtained with the basic sequence of models. 
Therefore, we will only report the results obtained with the basic
sequence of models.

Using the line fluxes I$^{\lambda}_{j}$ for individual  H\,{\sc ii} regions
normalized to I$^{H\beta}_{j}$, the global line fluxes I$^{\lambda}_{g}$
normalized to I$^{H\beta}_{g}$ can be determined.
The line fluxes R$_2$ $\equiv$ I$_{[OII]\lambda 3726 +\lambda 3729}$/I$_{H\beta}$,
R$_3$ $\equiv$ I$_{[OIII]\lambda 4959 +\lambda 5007}$/I$_{H\beta}$,  and
N$_2$ $\equiv$ I$_{[NII]\lambda 6548 +\lambda 6584}$/I$_{H\beta}$ are used in the present
study. The radial distribution of the number
density of H\,{\sc ii} regions changes significantly from galaxy to galaxy
(Rozas et al. 1996). The radial flux density distribution of all the
H\,{\sc ii} regions is very irregular. These variations of the flux
density are stronger than the systematic variations of the flux density with
galactocentric distance (Rozas et al. 1999). Therefore  the global oxygen and
nitrogen line fluxes I$^{*, \lambda}_{g}$ were derived as:
\begin{equation}
I_{g}^{*, \lambda}  =  \sum_{j=1}^{n}w_{j}I_{j}^{\lambda} ,
\label{equation:Igs}
\end{equation}
and the global H$_{\beta}$ line flux I$^{*, H_{\beta}}_{g}$  was derived as
\begin{equation}
I_{g}^{*, H_{\beta}}  =  \sum_{j=1}^{n}w_{j},
\label{equation:Ihb}
\end{equation}
where $n$ is the number of rings in a given model, and w$_j$ are random numbers
in the range 0 $\div$ 1. At a given $j$ the value w$_j$ is the same in
Eq.~\ref{equation:Igs} and Eq.~\ref{equation:Ihb}. We would like to stress that
the random distribution is not adopted for the luminosity function of
H\,{\sc ii} regions but for the radial flux distribution of all the
H\,{\sc ii} regions. The number of rings $n$ in a given
model is defined by the value $\Delta$Z = $Z_{\rm max}$ -- $Z_{\rm min}$
and corresponds to the number of H\,{\sc ii} regions with metallicity between
$Z_{\rm min}$ and $Z_{\rm max}$ in the selected subsample. Then the global
line fluxes I$^{\lambda}_{g}$ normalized to I$^{H\beta}_{g}$ are determined as:
\begin{equation}
I_{g}^{\lambda}  =  \frac{  \sum_{j=1}^{n}w_{j}I_{j}^{\lambda}} {  \sum_{j=1}^{n}w_{j}} .
\label{equation:Ig}
\end{equation}

The global line fluxes derived from Eq.~\ref{equation:Ig} are used for the
determination of global values of oxygen and nitrogen abundances.
The global oxygen and nitrogen abundances are derived within a two-zone model
for the temperature structure in the same way as in Paper I and Pilyugin, Thuan 
\& V\'{\i}lchez (2003). As a first step,
the (O/H)$_{P}$ oxygen abundance is determined with the expression suggested
in Pilyugin (2001):
\begin{equation}
12+log(O/H)_{P} = \frac{R_{23} + 54.2  + 59.45 P + 7.31 P^{2}}
                       {6.07  + 6.71 P + 0.37 P^{2} + 0.243 R_{23}}  ,
\label{equation:ohp}
\end{equation}
where $R_{\rm 23}$ =$R_{\rm 2}$ + $R_{\rm 3}$,
$R_{\rm 2}$ = $I_{[OII] \lambda 3727+ \lambda 3729} /I_{H\beta }$,
$R_{\rm 3}$ = $I_{[OIII] \lambda 4959+ \lambda 5007} /I_{H\beta }$,
and P = $R_{\rm 3}$/$R_{\rm 23}$.
Then, the electronic temperatures $T_{\rm e}$([OII]) and $T_{\rm e}$([OIII])
are derived using the value of O/H derived from Eq.\ref{equation:ohp} and
emission line fluxes derived from Eq.~\ref{equation:Ig}. For this purpose the
expressions for oxygen abundance determination from Pagel et al. (1992) and
the $T_{\rm e}$([OII]) -- $T_{\rm e}$([OIII]) relation from Garnett (1992) are used.
Then, assuming $t_{[NII]}=t_{[OII]}$, the N/O abundance ratio is determined
from the expression given in Pagel et al. (1992).

The 20 $\div$ 100 variants of random distributions of H\,{\sc ii} region 
H$_{\beta}$ luminosities have been considered for every model of galaxy with 
fixed values of $Z_{\rm min}$ and $Z_{\rm max}$. The oxygen abundances derived 
from the global spectra of the basic sequence of models are presented
in Fig.~\ref{figure:agoh} (filled circles) together with $Z_{\rm max}$
(open circles) and $Z_{\rm min}$ (open triangles) values.
The mean value of global oxygen abundances from a series of basic models, 
obtained by averaging the individual global oxygen abundances for different
variants of random distributions of  H\,{\sc ii} region H$_{\beta}$ luminosities,
versus oxygen abundance at galactocentric distances $r = 0.3R_{\rm 25}$
(asterisks), $r = 0.4R_{\rm 25}$ (open circles), and  $r = 0.5R_{\rm 25}$
(pluses) are presented in Fig.~\ref{figure:z04zg}. The oxygen abundance at a
given galactocentric distance is determined here as
12+log(O/H)$_{r=cR_{\rm 25}}$ = 12+log(O/H)$_{\rm max}$ -- c[log(O/H)$_{\rm max}$ --
log(O/H)$_{\rm min}$]), where $c=$0.3, 0.4, and 0.5. The data for the sequence
of models with $Z_{\rm min}$ = 8.42
are also presented in Fig.~\ref{figure:z04zg} by the filled circles.
The data for the sequence B of models (based on other subsample of H\,{\sc ii}
regions, see below) with $Z_{\rm min}$ = 8.30
are presented in Fig.~\ref{figure:z04zg} by the open squares. Examination of
Figs.\ref{figure:agoh} and \ref{figure:z04zg} shows that the oxygen abundance
derived from the global emission-line spectrum via the $P$--method agree
(within 0.1dex) with the oxygen abundance at
galactocentric distance $r=0.4R_{\rm 25}$, traced by individual H\,{\sc ii} regions,
if most of the H\,{\sc ii} regions belong to the upper branch of the O/H --
$R_{\rm 23}$ relation. It is worth noting that the oxygen abundances at
galactocentric distance $r=0.4R_{\rm 25}$ for the series B of the models are 
slightly higher than the global oxygen abundances.  The basic subsample of
H\,{\sc ii} regions corresponds to late-type spiral galaxies.
The subsample B of H\,{\sc ii} regions are close to the upper envelope in the
N/O versus O/H diagram (Fig.~\ref{figure:agno}) and corresponds to early-type spiral
galaxies.
Then, the global oxygen abundances derived for models of late-type galaxies 
are rather close to oxygen abundances at galactocentric distance $r=0.4R_{\rm 25}$ 
while the values derived for models of early-type galaxies are slightly lower.
The disagreement between the global oxygen abundances and the oxygen abundances
at galactocentric distance $r=0.4R_{\rm 25}$ for all the models does not
exceed 0.1 dex. This is in agreement with  the conclusion
of Kobulnicky, Kennicutt \& Pizagno (1999) that global emission-line spectra
can reliably indicate the chemical properties of galaxies.

The N/O ratios derived from global spectra of the basic serie of models are
presented in Fig.~\ref{figure:agno} (points) as a function of the global
oxygen abundance, together with the positions of individual H\,{\sc ii} regions
(open circles) from the basic subsample.
The N/O ratios derived from global spectra of the serie B of models are
presented by the filled squares,  the corresponding subsample of individual
H\,{\sc ii} regions  are shown by plus symbols.
Fig.~\ref{figure:agno} shows that the N/O ratios derived from global spectra
lie in the same region of the N/O -- O/H diagram as the individual H\,{\sc ii}
regions populating these galaxies.

The models of high-metallicity galaxies, i.e. models where all or the majority
of H\,{\sc ii} regions are metal-rich (12+log(O/H) $>$ 8.2) and lie on the
upper branch of the O/H -- $R_{\rm 23}$ diagram, have been considered
above. Now model spectra for the low-metallicity galaxies, populated by
metal-poor H\,{\sc ii} regions (12+log(O/H) $<$ 8.2) from the lower branch
of the O/H -- $R_{\rm 23}$ diagram, will be considered. It is well
known that the low-metallicity (irregular) galaxies show no significant
radial abundance gradient. Therefore the value of $\Delta$Z = $Z_{\rm max}$ --
$Z_{\rm min}$ $\sim$ 0.1dex is adopted. The H\,{\sc ii} regions from
the basic subsample are used. Again the 100 variants of random distributions 
of  H\,{\sc ii} region H$_{\beta}$ luminosities have been considered for each  
``artificial'' galaxy with given  $Z_{\rm min}$ and
$Z_{\rm max}$. The corresponding relation for low-metallicity H\,{\sc ii} regions
(Pilyugin 2001a,c) has been used for the oxygen abundance determination
instead of Eq.\ref{equation:ohp}.
The oxygen abundances in low-metallicity galaxies derived from the global
spectra are presented in Fig.~\ref{figure:agohlow} (points) together with the
$Z_{\rm max}$ (open circles) and $Z_{\rm min}$ (open triangles) values.
Inspection of Fig.~\ref{figure:agohlow} shows that the oxygen abundances
derived from the global emission-line spectra of a galaxy through the $P$--method 
are slightly lower than the oxygen abundances in individual H\,{\sc ii} regions.
It should be noted that the same effect has been revealed for the $T_{\rm e}$--method 
by Kobulnicky, Kennicutt \& Pizagno (1999). They  have found
that the oxygen abundances derived from the global emission-line spectra of
low-metallicity  galaxies via the $T_{\rm e}$--method are slightly underestimated,
by $\Delta$(O/H) $\le$ 0.1 dex.

The N/O ratios in low-metallicity galaxies derived from global spectra are
presented in Fig.~\ref{figure:agnolow} (points) as a function of the oxygen
abundance, together with the positions of individual H\,{\sc ii} regions
(open circles) from the basic subsample.
Fig.~\ref{figure:agnolow} shows that the N/O ratios derived from global spectra
of low-metallicity galaxies occupy the same band in the N/O -- O/H diagram as
the individual H\,{\sc ii} regions.

These results show that the oxygen abundances derived
from the global emission-line spectra of high-metallicity galaxies via 
the $P$--method agree within 0.1dex with the oxygen abundances at $r=0.4R_{\rm 25}$,
traced by individual H\,{\sc ii} regions,  if most of H\,{\sc ii} regions
belong to the upper branch of the O/H -- $R_{\rm 23}$ relation.
The oxygen abundances in low-metallicity galaxies derived from the global
emission-line spectra of galaxies via the $P$--method are slightly
underestimated by $\Delta$(O/H) $\le$ 0.1 dex.
The N/O ratios in high- and low-metallicity galaxies derived from global
spectra occupy the same band in the N/O -- O/H diagram as the individual
H\,{\sc ii} regions.
Our results confirm the conclusion of  Kobulnicky, Kennicutt \&
Pizagno (1999)  that the global emission-line spectra can reliably indicate
the chemical properties of galaxies.

\section{The chemical abundances in UV-selected galaxies}

\begin{table*}
\caption[]{\label{table:abundance}
Oxygen abundances and nitrogen-to-oxygen abundance ratios for the sample of
UV-selected galaxies from Contini et al. (2002).}
{\scriptsize
\begin{center}
\begin{tabular}{rcccccc}  \\ \hline \hline
       &             &             &            &             &            &             \\
Galaxy &  12+log(O/H)  & log(N/O)    & 12+log(O/H)  &   log(N/O)    & 12+log(O/H)  &   log(N/O)    \\
Number &   Original  & Original    & D--method  & D--method   & P--method  & P--method   \\
       &   Data      & Data        &            &             &            &             \\     
       &             &             &            &             &            &             \\   \hline
       &             &             &            &             &            &             \\
     1 &        8.99 &       -1.04 &       8.64 &       -0.96 &       8.67 &       -1.01 \\
     2 &        8.72 &       -1.58 &       8.34 &       -1.48 &       8.38 &       -1.54 \\
     3 &        8.72 &       -1.05 &       8.54 &       -0.85 &       8.38 &       -0.98 \\
     4 &        8.39 &       -0.93 &       7.97 &       -0.98 &            &             \\
     5 &        8.13 &       -1.83 &       7.95 &       -1.53 &            &             \\
     6 &        8.66 &       -1.13 &       8.39 &       -1.00 &       8.25 &       -1.04 \\
     7 &        8.20 &       -1.57 &       8.17 &       -1.52 &            &             \\
     8 &        8.31 &       -1.84 &       7.95 &       -1.53 &            &             \\
     9 &        8.78 &       -1.48 &       8.51 &       -1.26 &       8.39 &       -1.40 \\
    10 &        8.22 &       -1.62 &       8.23 &       -1.61 &            &             \\
       &             &             &            &             &            &             \\
    11 &        8.02 &       -1.29 &       8.48 &       -1.39 &       8.47 &       -1.40 \\
    12 &        8.38 &       -1.60 &       8.24 &       -1.43 &            &             \\
    13 &        8.90 &       -0.83 &       8.60 &       -0.78 &       8.59 &       -0.80 \\
    14 &        8.45 &       -1.18 &       8.13 &       -1.07 &            &             \\
    15 &        8.69 &       -1.23 &       8.51 &       -1.24 &       8.47 &       -1.28 \\
    16 &        8.80 &       -1.08 &       8.51 &       -1.10 &       8.50 &       -1.06 \\
    17 &        8.35 &       -1.74 &       8.16 &       -1.65 &            &             \\
    18 &        8.05 &       -1.73 &       8.27 &       -1.62 &            &             \\
    19 &        8.63 &       -1.22 &       8.33 &       -1.15 &       8.31 &       -1.20 \\
    20 &        8.66 &       -1.12 &       8.35 &       -1.09 &       8.34 &       -1.10 \\
       &             &             &            &             &            &             \\
    21 &        8.49 &       -1.26 &       8.17 &       -1.26 &            &             \\
    22 &        8.72 &       -1.20 &       8.33 &       -1.08 &       8.35 &       -1.14 \\
    23 &        8.65 &       -1.15 &       8.36 &       -1.13 &       8.36 &       -1.14 \\
    24 &        8.50 &       -0.91 &       8.16 &       -0.93 &            &             \\
    25 &        8.45 &       -1.01 &       8.29 &       -1.09 &            &             \\
    26 &        8.36 &       -1.51 &       8.07 &       -1.50 &            &             \\
    27 &        8.62 &       -1.05 &       8.34 &       -1.01 &       8.34 &       -1.04 \\
    28 &        8.42 &       -1.25 &       8.20 &       -1.12 &            &             \\
    29 &        8.73 &       -0.86 &       8.40 &       -0.90 &       8.41 &       -0.84 \\
    30 &        7.86 &       -1.78 &       7.95 &       -1.53 &            &             \\
       &             &             &            &             &            &             \\
    31 &        8.08 &       -1.79 &       8.27 &       -1.62 &            &             \\
    32 &        8.06 &       -1.71 &       8.27 &       -1.62 &            &             \\
    33 &        8.36 &       -1.44 &       8.05 &       -1.33 &            &             \\
    34 &        8.36 &       -1.52 &       8.07 &       -1.50 &            &             \\
    35 &        8.35 &       -1.67 &       8.02 &       -1.63 &            &             \\
    36 &        8.84 &       -1.25 &       8.56 &       -1.07 &       8.45 &       -1.18 \\
    37 &        8.50 &       -1.20 &       8.13 &       -1.07 &            &             \\
    38 &        8.63 &       -1.32 &       8.24 &       -1.25 &            &             \\
    39 &        8.36 &       -1.54 &       8.10 &       -1.50 &            &             \\
    40 &        8.61 &       -1.35 &       8.23 &       -1.28 &            &             \\
       &             &             &            &             &            &             \\
    41 &        8.37 &       -1.53 &       8.10 &       -1.50 &            &             \\
    42 &        8.77 &       -1.19 &       8.51 &       -1.15 &       8.51 &       -1.21 \\
    43 &        8.35 &       -1.65 &       8.02 &       -1.63 &            &             \\
    44 &        8.15 &       -1.74 &       7.95 &       -1.53 &            &             \\
    45 &        8.38 &       -1.47 &       8.11 &       -1.35 &            &             \\
    46 &        7.99 &       -1.82 &       8.35 &       -1.63 &       8.41 &       -1.86 \\
    47 &        8.46 &       -1.19 &       8.39 &       -1.25 &       8.36 &       -1.30 \\
    48 &        8.50 &       -1.22 &       8.24 &       -1.20 &            &             \\
    49 &        8.38 &       -1.48 &       8.11 &       -1.35 &            &             \\
    50 &        8.78 &       -1.10 &       8.44 &       -1.00 &       8.40 &       -1.03 \\
       &             &             &            &             &            &             \\
    51 &        8.72 &       -1.16 &       8.33 &       -1.08 &       8.38 &       -1.12 \\
    52 &        7.86 &       -1.49 &       8.53 &       -1.46 &       8.51 &       -1.55 \\
    53 &        7.78 &       -1.34 &       8.53 &       -1.46 &       8.61 &       -1.45 \\
    54 &        8.68 &       -1.13 &       8.39 &       -1.00 &       8.28 &       -1.05 \\
    55 &        8.38 &       -1.54 &       8.24 &       -1.43 &            &             \\
    56 &        8.73 &       -1.29 &       8.38 &       -1.23 &       8.42 &       -1.27 \\
    57 &        8.67 &       -1.23 &       8.39 &       -1.00 &       8.25 &       -1.14 \\
    58 &        8.90 &       -1.29 &       8.56 &       -1.07 &       8.53 &       -1.23 \\
    59 &        8.85 &       -1.11 &       8.56 &       -1.07 &       8.48 &       -1.06 \\
    60 &        8.05 &       -1.53 &       8.34 &       -1.52 &       8.34 &       -1.54 \\
       &             &             &            &             &            &             \\
    61 &        8.03 &       -1.75 &       7.95 &       -1.53 &            &             \\
    62 &        8.94 &       -1.17 &       8.63 &       -1.10 &       8.65 &       -1.16 \\
    63 &        8.72 &       -1.19 &       8.39 &       -1.00 &       8.33 &       -1.12 \\
    64 &        8.82 &       -1.36 &       8.51 &       -1.26 &       8.47 &       -1.33 \\
    65 &        8.73 &       -1.40 &       8.34 &       -1.48 &       8.41 &       -1.38 \\
    66 &        8.36 &       -1.50 &       8.05 &       -1.33 &            &             \\
    67 &        8.91 &       -1.08 &       8.63 &       -1.10 &       8.60 &       -1.06 \\
    68 &        8.81 &       -1.28 &       8.51 &       -1.26 &       8.43 &       -1.22 \\
       &             &             &            &             &            &             \\
   \\  \hline
\end{tabular}
\end{center}
}
\end{table*}

\begin{figure}
\resizebox{1.00\hsize}{!}{\includegraphics[angle=270]{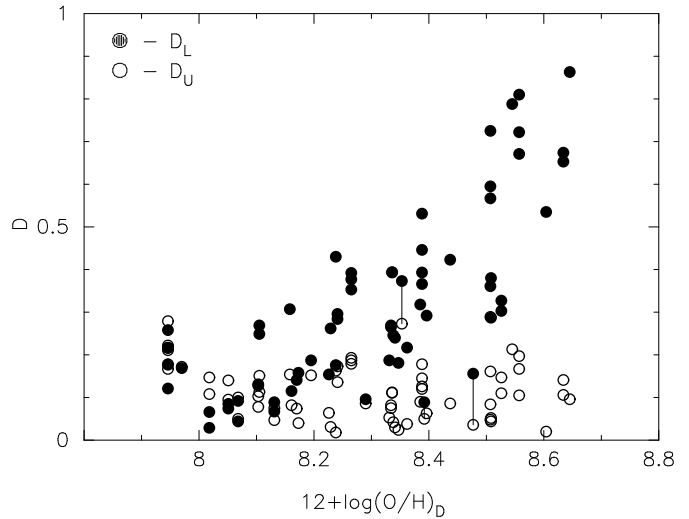}}
\caption{The difference indexes D$_U$ (open circles) and D$_L$ (filled circles)
as a function of the 12+$\log$(O/H)$_D$ for the sample of UV-selected
galaxies (Contini et al. 2002).
}
\label{figure:ohdd}
\end{figure}

\begin{figure}
\resizebox{1.00\hsize}{!}{\includegraphics[angle=000]{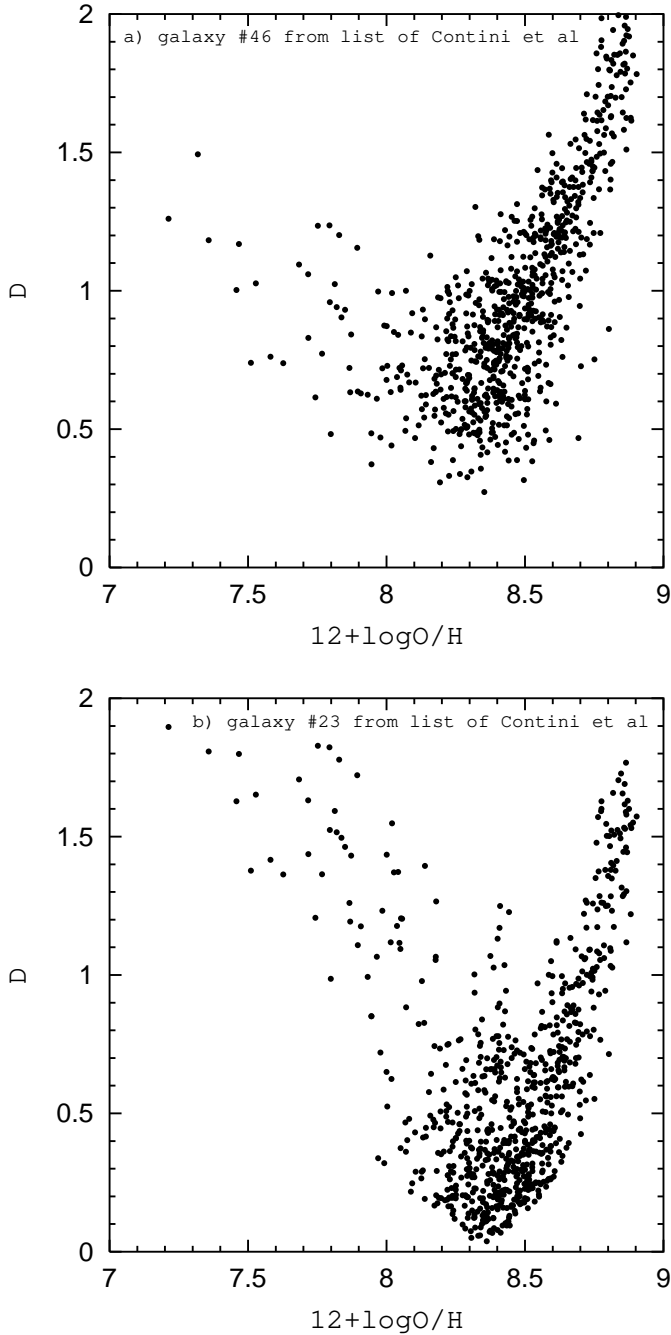}}
\caption{The difference indexes D$_j$ versus the oxygen abundances in
individual H\,{\sc ii} regions. Panel {\bf a)} shows the data for the galaxy
\#46 ("peculiar" case) from the list of Contini et al. (2002).
For comparison, panel {\bf b)} shows the data for galaxy \#23 ("normal"
case) from the list of Contini et al. (2002).
}
\label{figure:wk9}
\end{figure}

\begin{figure}
\resizebox{1.00\hsize}{!}{\includegraphics[angle=270]{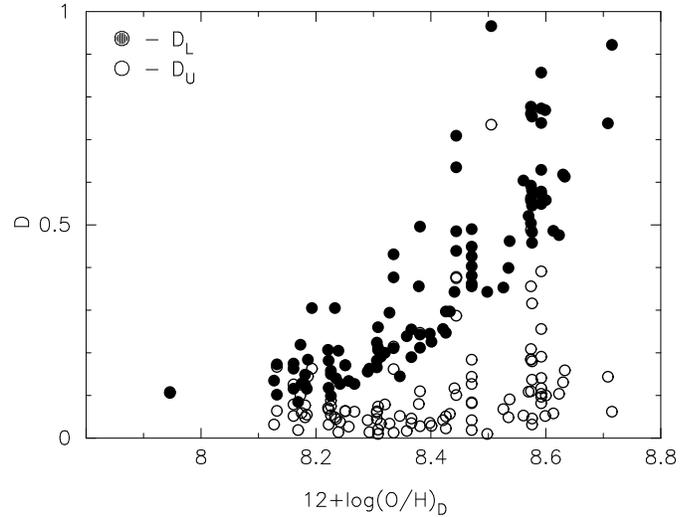}}
\caption{The difference indexes D$_U$ (open circles) and D$_L$ (filled circles)
as a function of the 12+$\log$(O/H)$_D$ for the sample of nearby
galaxies (Jansen et al. 2000).
}
\label{figure:ohddj}
\end{figure}

\begin{figure}
\resizebox{0.95\hsize}{!}{\includegraphics[angle=270]{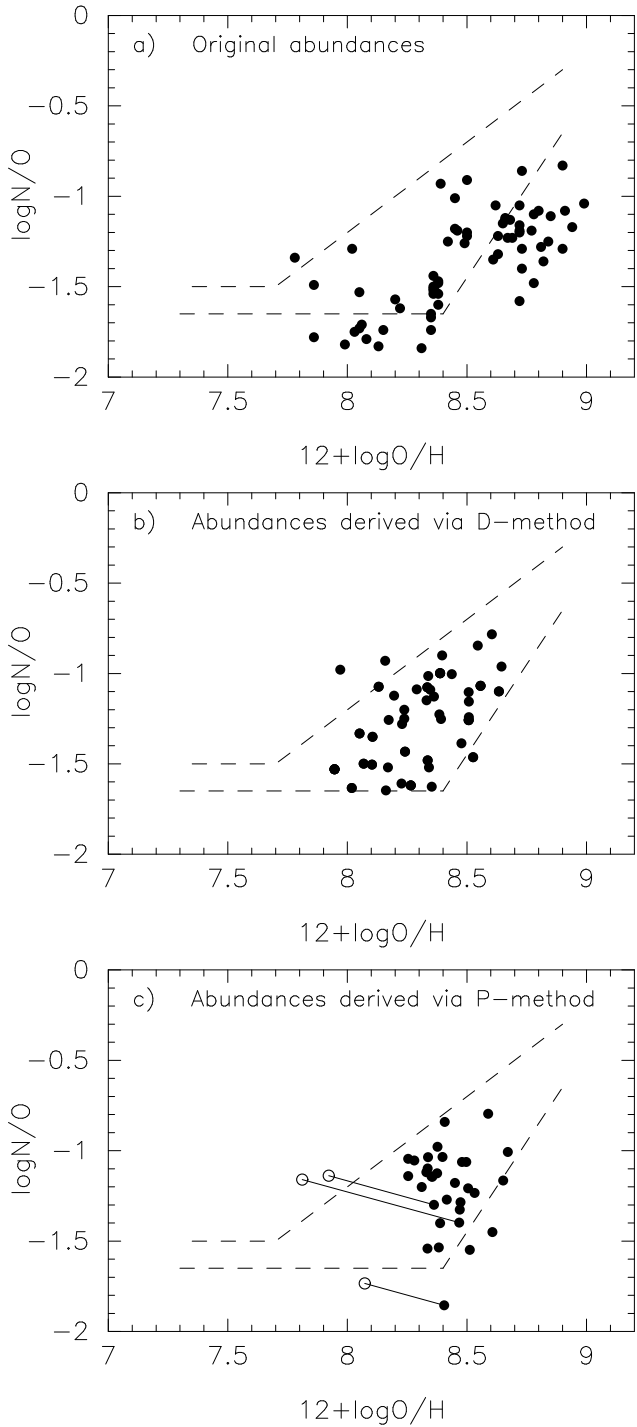}}
\caption{The N/O vs. O/H diagram for the sample of UV-selected galaxies from
Contini et al. (2002). The dashed lines outline the area occupied
by individual H\,{\sc ii} regions from our collection (see Fig.~\ref{figure:fxxx}).
{\bf a)} Filled circles are original abundances from Contini et al. (2002)
determined via the calibration of Kobulnicky, Kennicutt \& Pizagno (1999).
{\bf b)} Filled circles are abundances estimated using the $D$--method (see text).
{\bf c)} Filled circles are abundances in galaxies with
12+log(O/H)$_{D}$ $>$ 8.3 determined via the high-metallicity $P$--calibration.
The open circles are abundances determined through the low-metallicity $P$--calibration
in three galaxies with unreliable classification. The abundances derived via the
high- and  low-metallicity $P$--calibrations for the same galaxy are connected
with a solid line.
}
\label{figure:nooh}
\end{figure}

\begin{figure}
\resizebox{1.00\hsize}{!}{\includegraphics[angle=270]{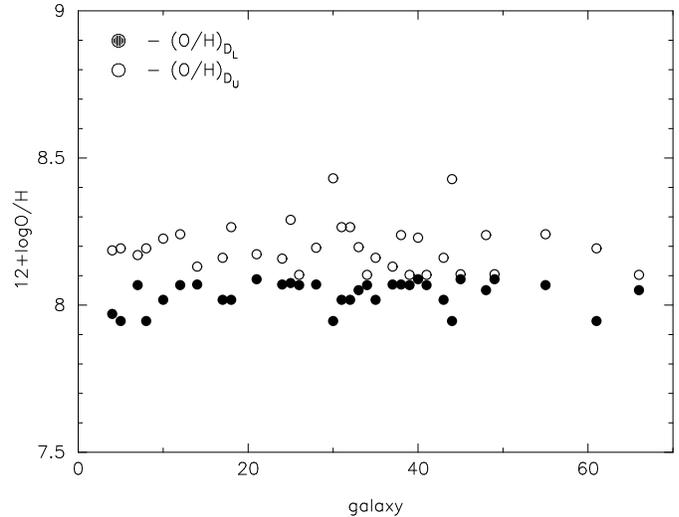}}
\caption{The oxygen abundances in galaxies with 12+log(O/H)$_D$ $<$ 8.3.
The open circles are the oxygen abundances
derived through the $D$--method with the D$_U$ values, i.e. under the assumption
that the galaxy has an oxygen abundance higher than 12+log(O/H) = 8.1. The value
of D$_{\rm min}$ = D$_U$ is determined by comparing the spectra of galaxies with
the spectra of H\,{\sc ii} regions with 12+log(O/H) $>$ 8.1.
The filled circles are the oxygen abundances
derived through the $D$--method with the D$_L$ value, i.e. under the assumption
that the galaxy has an oxygen abundance lower than 12+log(O/H) = 8.1 and the value
of D$_{\rm min}$ = D$_L$ is determined by comparing the spectra of galaxies with
the spectra of H\,{\sc ii} regions with 12+log(O/H) $<$ 8.1.
}
\label{figure:ohlohu}
\end{figure}

We have shown that the P--method can be used to derive chemical abundances from
the global emission-line spectra of galaxies. As in case of individual
H\,{\sc ii} regions, one has to know apriori on which of the two branches the
galaxy lies.
We will try to solve this problem with the following strategy. We will compare
the global spectrum of a galaxy with the spectrum of a large number of
individual H\,{\sc ii} regions with known oxygen abundances. The oxygen abundance of
a H\,{\sc ii} region, which shows the minimum spectral difference with the global
spectrum of a galaxy, can be adopted as the global oxygen abundance of this galaxy.
The difference between the global spectrum of a galaxy and the spectrum of an individual
H\,{\sc ii} region can be specified by the parameter D:
\begin{eqnarray}
D  =  [(\log R_{2}^{g} - \log R_{2}^{HII})^{2} +
(\log R_{3}^{g} - \log R_{3}^{HII})^{2}  \nonumber \\
+ (\log N_{2}^{g} - \log N_{2}^{HII})^{2}]^{\frac {1}{2}}
\label{equation:d}
\end{eqnarray}
where R$_{2}^{g}$, R$_3^{g}$, and N$_2^{g}$ are the line fluxes in global
spectra of galaxies normalized to the H$_{\beta}$ flux. R$_{2}^{HII}$,
R$_3^{HII}$, and N$_2^{HII}$ are the line fluxes in spectra of individual
H\,{\sc ii} regions  normalized to the H$_{\beta}$ flux. The use of the
logarithms instead of the fluxes in the definition of $D$ is due to the
following reason. Since the [OII], [OIII], and [NII] line fluxes can be
significantly different, the $D$ parameter would be dominated by the
strongest emission lines if fluxes were used in the definition.

The comparison of global spectra of 68 UV-selected galaxies from Contini et al.
(2002) with spectra of individual H\,{\sc ii} regions was carried out
using our collection of 785 individual H\,{\sc ii} regions.
The 785 values of the difference index D$_j$ have been determined from
Eq.~\ref{equation:d} for each  UV-selected galaxy from the sample of Contini
et al. (2002). We choose from D$_j$ the minimum value of difference index
$D_{\rm min}$, and the oxygen abundance in the galaxy is adopted to be equal to the
oxygen abundance in the corresponding H\,{\sc ii} region. The oxygen abundance
in the galaxy estimated in such a way will be referred to (O/H)$_D$.
Then we choose from D$_j$ the minimum value of difference index
D$_L$ when the spectrum of the galaxy is compared to the spectra of
H\,{\sc ii} regions with 12+log(O/H) $\le$ 8.1, and the minimum value of difference
index D$_U$ when the spectrum of the galaxy is compared to the spectra of
H\,{\sc ii} regions with 12+log(O/H) $>$ 8.1. It is evident that the
value of $D_{\rm min}$ is equal to the value of D$_L$ or to the value of D$_U$.
Fig.~\ref{figure:ohdd} shows the value of D$_U$ (open circles) and the values of
D$_L$ (filled circles) as a function of 12+log(O/H)$_D$ for the sample of UV-selected
galaxies.

Inspection of Fig.~\ref{figure:ohdd} shows that it is possible to find, among
the spectra of individual H\,{\sc ii} regions, a spectrum quite similar to the
global spectrum of each UV-selected galaxy. It also shows that, for a fraction
of UV-selected galaxies, a similar spectrum can be found among the
spectra of both high-metallicity and low-metallicity H\,{\sc ii} regions, the values of
 D$_U$ and D$_L$ being relatively close to each other. But for a significant fraction
of UV-selected galaxies, a similar spectrum can be found only among the
spectra of high-metallicity H\,{\sc ii} regions, the D$_U$ value being significantly
lower than the value of D$_L$. The value of D$_U$ is appeciably lower than the
value of D$_L$ for galaxies with 12+log(O/H)$_D$ $>$ 8.3 (Fig.~\ref{figure:ohdd}).
Then, the condition 12+log(O/H)$_D$ $>$ 8.3 (or D$_U$ $<$ D$_L$) allows us to
identify the high-metallicity galaxies.
Fig.~\ref{figure:ohdd} shows that the application of these criteria
is not beyond question for three cases. Two galaxies, \#11 and \#47,
have low values both for D$_U$ and D$_L$. On the contrary, both the value of
D$_U$ and the value of D$_L$ are high for the galaxy \#46 from the list of
Contini et al. (2002). The values of D$_U$ and D$_L$ for
these galaxies are connected by solid lines in Fig.~\ref{figure:ohdd}.
The values of the difference index D$_j$ for the galaxy \#46 ("peculiar" case)
versus the oxygen abundances in individual H\,{\sc ii} regions are shown in 
Fig.~\ref{figure:wk9} (panel {\bf a}). For comparison, the values of the
difference index D$_j$ for a "normal" case (galaxy \#23) of similar metallicity 
versus the oxygen abundances in individual H\,{\sc ii} regions are shown in 
Fig.~\ref{figure:wk9} (panel {\bf b}).
Thus, the condition  12+log(O/H)$_D$ $>$ 8.3 (or D$_U$ $<$ D$_L$) allows us to
identify the high-metallicity galaxies. The confidence level of the identification
is defined by the values of D$_U$ and D$_L$. If the value of D$_U$ is low and
the value of D$_L$ is significantly higher than the value of
D$_U$ the high-metallicity galaxies are identified with a high probability.

Thus, the comparison of the global spectrum of a galaxy with a collection of spectra
of individual H\,{\sc ii} regions can be used to distinguish
high- versus low-metallicity objects.
To verify this conclusion, the sample of nearby field galaxies
from Jansen et al. (2000) has been considered.
Fig.~\ref{figure:ohddj} shows the values of D$_U$ (open circles) and D$_L$
(filled circles) as a function of 12+log(O/H)$_D$ for the sample 
of Jansen et al. (2000)\footnote{Jansen et al. (2000) published both
the integrated and nuclear spectra of galaxies. For this study, we use only the
integrated spectra.}. Comparison of Fig.~\ref{figure:ohdd} and
Fig.~\ref{figure:ohddj} shows that the trends of values for D$_U$ and
D$_L$ as a function of 12+log(O/H)$_D$ are quite similar for both samples of galaxies.

Using the conditions listed above, we selected the galaxies from
Contini et al. (2002) which are expected to lie on the upper branch of the
O/H -- $R_{\rm 23}$ relation. Chemical abundances in these galaxies were
determined through the high-metallicity $P$--calibration.
The derived abundances are presented in Table \ref{table:abundance}.
The galaxy number according to the list of Contini et al. (2002) is
listed in column 1.
The original oxygen abundance and nitrogen-to-oxygen abundance
ratio are given in columns 2 and 3.
The oxygen abundance and nitrogen-to-oxygen abundance ratio estimated via the 
$D$--method are reported in columns 4 and 5.
The oxygen abundance and nitrogen-to-oxygen abundance ratio determined through
the $P$--method for high-metallicity galaxies are listed in columns 6 and 7.

It should be stressed
that the lower limit of oxygen abundances where the high-metallicity
$P$--calibration is suitable for spiral galaxies does not coincide with that for
individual H\,{\sc ii} regions. Examination of Figs.\ref{figure:agoh} and
\ref{figure:z04zg} shows that the $P$--calibration for the high-metallicity
branch of the O/H -- $R_{\rm 23}$ relation provides realistic global oxygen
abundances only for
galaxies with a global oxygen abundance higher than 12+log(O/H) $\sim$ 8.3,
while the high-metallicity $P$--calibration provides realistic
oxygen abundances in individual H\,{\sc ii} regions with 12+log(O/H) $>$ 8.2.
It is due to the following reason. Spiral galaxies show usually a
radial abundance gradient. If a spiral galaxy has a global oxygen abundance
in the range 8.2 $<$ 12+log(O/H) $<$ 8.3 then this spiral is populated both by H\,{\sc ii}
regions with 12+log(O/H) $>$ 8.2 and by H\,{\sc ii} regions with 12+log(O/H)
$<$ 8.2. H\,{\sc ii} regions with 12+log(O/H) $<$ 8.2 can make a
significant contribution to the global spectrum of such a galaxy. As a result,
the high-metallicity $P$--calibration cannot be used for abundance determination
in this type of galaxy. Formally, one can say that the boundary between the upper
branch and the transition zone in the O/H -- $R_{\rm 23}$ relation
for spiral galaxies does not coincide with that for individual H\,{\sc ii} regions.

The N/O -- O/H diagram for the UV-selected galaxies is presented in
Fig.~\ref{figure:nooh}. Panel {\bf a)} shows the original oxygen abundances
and nitrogen-to-oxygen abundance ratios from Contini et al. (2002)
determined with the calibration of Kobulnicky, Kennicutt \& Pizagno (1999).
The dashed lines outline the area occupied by the individual
H\,{\sc ii} regions from our collection (see Fig.~\ref{figure:fxxx}).
The oxygen abundances and nitrogen-to-oxygen abundance
ratios estimated in UV-selected galaxies via the D--method  are
presented in panel {\bf b)}. Panel {\bf c)} shows the oxygen abundances and
nitrogen-to-oxygen abundance ratios determined through the P--method for
high-metallicity galaxies selected according to the criteria listed above.
The comparison of panel {\bf a)} and panel {\bf c)} of Fig.~\ref{figure:nooh}
shows that the N/O -- O/H diagram for high-metallicity (12+log(O/H) $>$ 8.3)
UV-selected galaxies based on the
abundances determined through the P--method  differs significantly from that 
based on the original abundances derived by Contini et al. (2002). The UV-selected
galaxies with abundances determined using the P--method occupy the same
area as the individual H\,{\sc ii} regions in normal spiral galaxies. The
shift of original positions of UV-selected galaxies could thus be
due to calibration problems for abundance determination.

How many high-metallicity (12+log(O/H) $>$ 8.3) galaxies are lost using the
selection criteria listed above? Indeed, the typical value of D$_U$ for
low-metallicity galaxies is close to that for high-metallicity
galaxies (see Fig.~\ref{figure:ohdd}). Then, one can
expect that the high-metallicity galaxies can exist among the galaxies classified
as low-metallicity ones, with 12 + log(O/H) $<$ 8.3. Fig.~\ref{figure:ohlohu}
shows the two values of oxygen abundance in galaxies classified as
low-metallicity.  The open circles correspond to the oxygen abundances
derived through the D--method with the D$_U$ value, i.e. under the assumption
that the galaxy has an oxygen abundance higher than 12+log(O/H) = 8.1 and the value
of $D_{\rm min}$ = D$_U$ is determined by comparing the spectra of galaxies with
the spectra of H\,{\sc ii} regions with 12 + log(O/H) $>$ 8.1.
The filled circles correspond to the oxygen abundances
derived through the D--method with the D$_L$ value, i.e. under the assumption
that the galaxy has an oxygen abundance lower than 12+log(O/H) = 8.1 and the value
of $D_{\rm min}$ = D$_L$ is determined by comparing the spectra of galaxies with
the spectra of H\,{\sc ii} regions with 12 + log(O/H) $<$ 8.1.
Examination  of Fig.~\ref{figure:ohlohu} shows that both oxygen abundances
12+log(O/H)$_L$ and oxygen abundances 12+log(O/H)$_U$ in galaxies classified
as low-metallicity ones are below 12+log(O/H) = 8.3, with two
exceptions. The classification of only two galaxies, \#30 and \#44, may be wrong.
The bulk of galaxies classified as low-metallicity ones appear
to belong to the transition zone of the O/H -- $R_{23}$ relation, or some of them
lie on the lower branch of this relation. Thus, the selection criterion used in this 
study seems to be reliable.

The typical difference between oxygen abundance determined in high-metallicity
galaxies via the P--method and that estimated via the D--method
is small, i.e. $<$ 0.1 dex (see Table \ref{table:abundance}). Since the values of
$D_{\rm min}$ for low-metallicity galaxies are similar to those for high-metallicity
galaxies (see Fig.~\ref{figure:ohdd}), one can hope that the D--method also 
provides realistic estimations of the oxygen
abundance in low-metallicity galaxies which belong to the transition zone
or lie on the lower branch of the O/H -- $R_{\rm 23}$ relation.
It should be noted that the oxygen abundance in our collection of
individual H\,{\sc ii} regions with 12+log(O/H) $<$ 8.2 was 
determined with the T$_e$--method. Panel {\bf b)}
of Fig.~\ref{figure:nooh} shows that the low-metallicity UV-selected galaxies
fill more or less uniformily the area outlined by the individual H\,{\sc ii}
regions in the N/O -- O/H diagram, i.e. the low-metallicity UV-selected
galaxies do not  show any shift relative to the positions of individual
H\,{\sc ii} regions in local spiral galaxies.

The fact that the typical difference between O/H determined with 
the P--method and that estimated via the D--method is small
(see Table~\ref{table:abundance}) shows that our
H\,{\sc ii} region spectral database is large enough so that any spectrum of
UV-selected galaxies considered in this study can be represented by one of
these H\,{\sc ii} region spectra.

\section{The luminosity -- metallicity relation}

\begin{figure}
\resizebox{1.00\hsize}{!}{\includegraphics[angle=270]{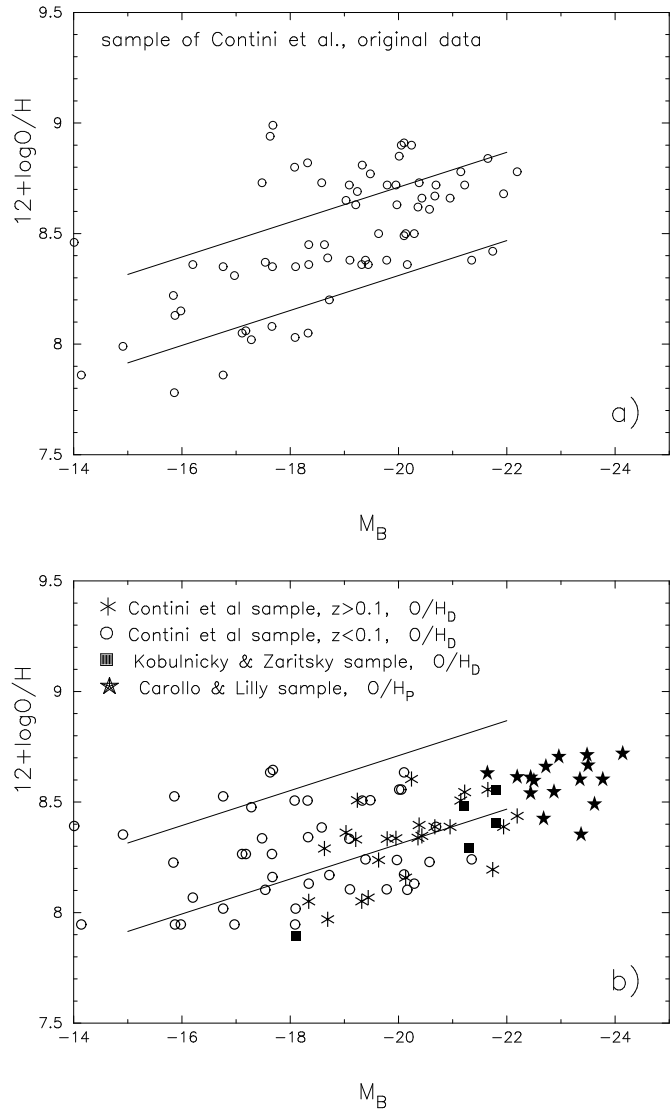}}
\caption{Luminosity -- metallicity relation.
{\bf a)} Open circles are the original data for the
sample of UV-selected local and intermediate-redshift galaxies from Contini
et al. (2002). The solid lines correspond to the "luminosity -- metallicity band"
occupied by well-studied local spiral galaxies (Paper I).
{\bf b)} The luminosity -- metallicity relation for the
the Contini et al's sample of UV-selected local ($z < 0.1$; open circles)
and intermediate-redshift ($z > 0.1$; asterisks) galaxies
with oxygen abundances re-determined via the $D$--method.
The filled squares are the intermediate-redshift ($0.1 < z < 0.5$) galaxies from
the sample of Kobulnicky \& Zaritsky (1999) with oxygen abundances re-derived
via the $D$--method and absolute blue magnitudes scaled to the
value of H$_0$ = 100 km s$^{-1}$ Mpc$^{-1}$. The stars are
intermediate-redshift ($0.5 < z < 1.0$) galaxies from the sample of Carollo \& Lilly (2001) with
oxygen abundances derived via the $P$--method and absolute blue
magnitudes scaled to the value of H$_0$ = 100 km s$^{-1}$ Mpc$^{-1}$.
}
\label{figure:lz}
\end{figure}

\begin{figure}
\resizebox{1.00\hsize}{!}{\includegraphics[angle=270]{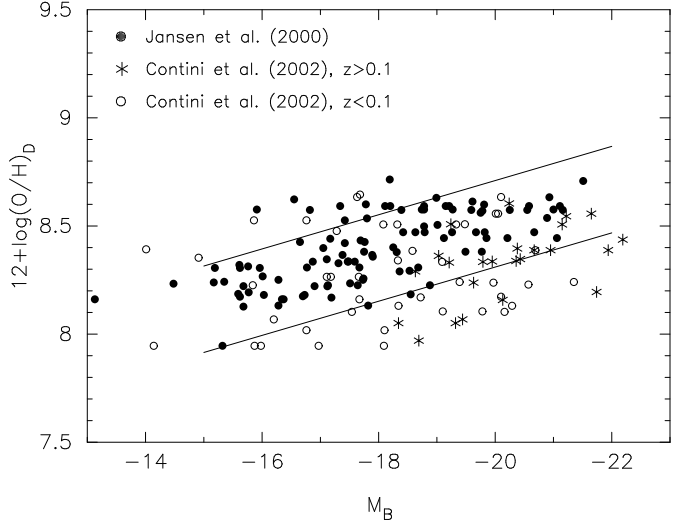}}
\caption{Luminosity -- metallicity relations for the sample of nearby
galaxies (filled circles) from Jansen et al. (2000) and for the sample
of UV-selected local ($z < 0.1$; open circles) and intermediate-redshift
($z > 0.1$; asterisks)
galaxies from Contini et al. (2002).
The solid lines show the "luminosity -- metallicity band"
obtained in Paper I for well-studied local spirals.
}
\label{figure:lzjc}
\end{figure}

\begin{figure}
\resizebox{1.00\hsize}{!}{\includegraphics[angle=270]{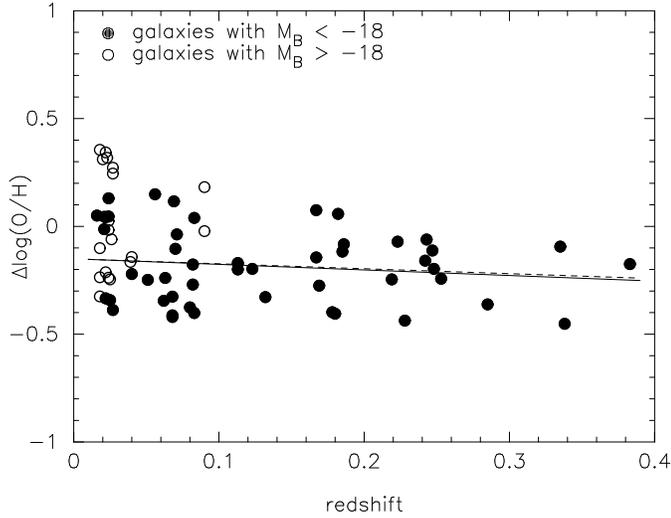}}
\caption{The deviation of the oxygen abundance from the general luminosity --
metallicity trend (obtained in Paper I) as a function of redshift for the
sample of  UV-selected galaxies from Contini et al. (2002).
The filled circles are galaxies with $M_{\rm B} < -18$,
the open  circles are galaxies with $M_{\rm B} > -18$. The solid line corresponds
to the best fit to the data for galaxies with $M_{\rm B}  < -18$, the dashed
line corresponds to the best fit to galaxies with redshift $z > 0.1$.
}
\label{figure:zdoh}
\end{figure}

The luminosity -- metallicity relation for UV-selected local and
intermediate-redshift (0 $<$ z $\le$ 0.4) galaxies of Contini et al (2002)
is presented in Fig.~\ref{figure:lz}. The upper panel {\bf a)} shows the original data.
The luminosity -- metallicity relation with oxygen abundances re-determined via
the $D$--method is shown in the lower panel.
The global optical spectra for 14 star-forming emission-line galaxies at
intermediate redshifts ($0.1 < z < 0.5$) have been published by Kobulnicky \&
Zaritsky (1999).
The oxygen abundances in five galaxies with measured line fluxes
I$_{[OII]\lambda 3726 +\lambda 3729}$,  I$_{[OIII]\lambda 4959 +\lambda 5007}$,
and I$_{[NII]\lambda 6584}$ (L2-408115, L2-410083, SA 68-206134,
SDG 223, and SA 68-207213) from the sample of Kobulnicky \& Zaritsky (1999)
have been estimated here via the $D$--method. The position of these
galaxies in the luminosity -- metallicity diagram is also shown (filled squares)
in the lower panel of Fig.~\ref{figure:lz}. To compare the luminosity
-- metallicity relation for the samples of galaxies from Kobulnicky \& Zaritsky
(1999) and from Contini et al. (2002), the absolute magnitudes of galaxies from
Kobulnicky \& Zaritsky (1999) have been decreased by $-1.5$ mag since they used
the value of H$_0$ = 50 km s$^{-1}$ Mpc$^{-1}$ while the absolute magnitude for
galaxies of Contini et al. were estimated with H$_0$ = 100 km s$^{-1}$ Mpc$^{-1}$.
The solid lines in Fig.~\ref{figure:lz} outline the "luminosity -- metallicity
band" occupied by well-studied local spiral galaxies (Paper I).

Fig.~\ref{figure:lz}b shows that the UV-selected
intermediate-redshift sample ($z > 0.1$), as well as the sample of intermediate-redshift
galaxies from Kobulnicky \& Zaritsky (1999), have, on average, lower oxygen abundances 
compared to the local galaxies of the same luminosity or/and that the intermediate-redshift
galaxies are slightly more luminous than local galaxies of the same metallicity. 
This does not agree with the conclusions of Carollo \& Lilly (2001). 
They measured the emission-line flux
ratios in global spectra of intermediate-redshift ($0.5 < z < 1.0$) galaxies. Based on the
similarity between the positions of these galaxies and that of local
field galaxies from the sample of Jansen et al (2000) in the $R_{\rm 23}$ --
[OIII]/[OII] and the $R_{\rm 23}$ -- $M_{\rm B}$ diagrams, they concluded that
the metallicities of the intermediate-redshift galaxies appear to be remarkably
similar to those of local field galaxies, and there appears to have been little
change in the relationship between metallicity and luminosity from $z \sim 1$
to today. It should be noted that the absolute blue magnitudes of galaxies in
these samples are estimated with different values of H$_0$;
with H$_0$ = 50 km s$^{-1}$ Mpc$^{-1}$ in the sample of Carollo \& Lilly (2001)
and with H$_0$ = 100 km s$^{-1}$ Mpc$^{-1}$ in the sample of Jansen et al (2000).
Then the direct comparison of the $R_{\rm 23}$ -- $M_{\rm B}$ diagram for these samples
of galaxies is not justified.
The oxygen abundances in 13 galaxies (two AGNs were excluded
from our consideration) from the sample of Carollo \& Lilly (2001)
with measured line fluxes as well as in 3 galaxies with
estimated upper limits of [OIII]$\lambda$$\lambda$4959,5007 line fluxes
were re-determined using the P--method\footnote{The $D$--method cannot
be used for the abundance determination in Carollo \& Lilly's galaxies since
the [NII]$\lambda 6584$ line measurements are not available.}.
The absolute magnitudes of galaxies from
Carollo \& Lilly (2001) have been decreased by $-1.5$ mag since they used
the value of H$_0$ = 50 km s$^{-1}$ Mpc$^{-1}$ while the absolute magnitude for
galaxies of Contini et al. have been estimated with H$_0$ = 100 km s$^{-1}$ Mpc$^{-1}$.
The O/H -- $M_{\rm B}$ diagram for the sample of Carollo \&
Lilly (2001) is presented in lower panel of Fig. \ref{figure:lz} by stars.
It is clear that the intermediate-redshift galaxies from Carollo \& Lilly
fall in the O/H versus $M_{\rm B}$ plane defined by other intermediate-redshift
galaxies and show a shift relative to the local field galaxy sample.

It should be noted however that the luminosity of intermediate-redshift
galaxies  has been estimated using the galaxy redshifts. As a consequence,
the luminosity of intermediate-$z$ galaxies depends on the adopted value of
the Hubble constant H$_0$, while accurate
distance determinations with high-precision methods (the
Cepheid period-luminosity relation, the peak brightness of type Ia supernovae,
the expanding photospheres of type II supernovae and others) are available
for the majority of local spiral galaxies (see Paper I).
The absolute blue luminosity of local galaxies does not depend on the value of
the Hubble constant H$_0$. Could the shift in the positions of the
intermediate-redshift galaxies in the luminosity -- metallicity diagram
relative to the positions of the local spiral galaxies be due to the choice
(somewhat arbitrary) of the Hubble constant value?

Jansen et al. (2000) published global spectra for a representative sample
of nearby galaxies.
The oxygen abundances in nearby galaxies with measured global line fluxes
I$_{[OII]\lambda 3726 +\lambda 3729}$,  I$_{[OIII]\lambda 4959 +\lambda 5007}$,
and I$_{[NII]\lambda 6548 + \lambda 6584}$ from the sample of Jansen et al.
(2000) were estimated via the D--method.
The luminosity -- metallicity diagram for the sample of nearby galaxies from Jansen
et al. (2000) is shown in Fig.~\ref{figure:lzjc}. Examination of
this figure shows that the local galaxies from the sample of
Jansen et al (2000) occupy the same band in the luminosity -- metallicity diagram
as local galaxies from the sample of Paper I.
The absolute blue magnitudes for these galaxies have been calculated by Jansen et al.
from the total apparent $B$-band magnitude and the galaxy redshifts, assuming a
simple Hubble flow and H$_0$ = 100 km s$^{-1}$ Mpc$^{-1}$. Therefore
the luminosity -- metallicity relation for the sample of Jansen et al. (2000)
can be directly compared to the
luminosity -- metallicity relation for the sample of Contini et al. (2002).

In Figure~\ref{figure:lzjc}, we compare the luminosity -- metallicity relations
for the sample of Contini et al. (2002) and Jansen et al. (2000). It shows
that a significant fraction of UV-selected galaxies are below the luminosity -- metallicity
defined by normal nearby galaxies.
Thus, the shift in the position of the intermediate-redshift galaxies in
the luminosity -- metallicity diagram relative to the position of the nearby
galaxies seems to be real and does not depend on the choice of the Hubble constant value.

It thus appears that the intermediate-redshift galaxies systematically deviate from the
metallicity -- luminosity trend of local galaxies. One interpretation could be
that the intermediate-redshift galaxies are slightly less advanced in their
chemical evolution and, as a consequence, are more metal-deficient than nearby galaxies.
Another interpretation could be that intermediate-$z$ galaxies are observed at a special
stage in their evolution; they have just undergone a powerful starburst which
temporary lowered their mass-to-light ratio, resulting in a temporary increase
of their blue luminosity.

If intermediate-redshift galaxies are slightly less advanced in their evolution,
one can expect that the deviation from the local $L-Z$ relation correlates with
redshift. The deviation of oxygen abundance from the local metallicity -- luminosity
relation was estimated in the following way.
Taking into account {\it i)} that the H$_{0}$-dependent luminosity --
metallicity relation for local galaxies (Jansen et al. 2000) is in good agreement
with the H$_{0}$-independent $L-Z$ relation for the well-studied local galaxies of
Paper I (see Fig.~\ref{figure:lzjc}, and {\it ii)} that absolute blue magnitudes for
UV-selected and nearby galaxies have been calculated
with the same value of H$_{0}$, the metallicity -- luminosity relationship
from Paper I,
\begin{equation}
12 + \log(O/H)   =  6.93 - 0.079 \, M_B ,
\label{equation:lz}
\end{equation}
can be used to estimate the deviation of oxygen abundance from the 
metallicity -- luminosity trend. Then, the value of the oxygen abundance
deviation is $\Delta$log(O/H) = (12+log(O/H)$_D$) -- (12+log(O/H)$^*$),
where 12+log(O/H)$^*$ is estimated from the absolute blue magnitude using
Eq.~\ref{equation:lz}.
The deviation of the oxygen abundance from the general luminosity --
metallicity trend as a function of redshift for the sample of  UV-selected
galaxies from Contini et al. (2002) is presented in Fig.~\ref{figure:zdoh}.
The filled circles are galaxies with $M_{\rm B}  < -18$, the open  circles are
galaxies with $M_{\rm B}  > -18$. Examination of Fig.~\ref{figure:zdoh}
shows that the dispersion of points increases with decreasing redshift and
reaches a maximum value near $z=0$. This could be due to the fact that the distance
and consequently the absolute blue magnitude for galaxies calculated from the
galaxy redshifts are less accurate at low redshifts since peculiar motions can make
a significant contribution to the observed values of $z$. On the other hand,
close examination of Fig.~\ref{figure:zdoh} shows that the low-luminosity
galaxies, with $M_{\rm B}  > -18$, have a larger dispersion than luminous ones.
It can indicate that the emission-line intensity measurements
are more accurate in luminous than in low-luminosity galaxies.

The solid line in Fig.~\ref{figure:zdoh} corresponds to the best fit to the data
for galaxies with $M_{\rm B} < -18$, the dashed
line corresponds to the best fit to galaxies with redshift $z > 0.1$.
Both relations show that the position of local ($z \sim 0$) UV-selected
galaxies are shifted from the luminosity-metallicity trend of normal galaxies.
It suggests that a temporary increase of their blue luminosity could contribute
significantly to the shift of UV-selected galaxies in the luminosity-metallicity
diagram.

Fig.~\ref{figure:zdoh} shows that there is a marginal correlation
(correlation coefficient r $<$ 0.1) between the
deviation of oxygen abundance in a galaxy from the metallicity -- luminosity
trend and galaxy redshift. If one tries to coincide the positions of UV-selected galaxies
at $z=0$ with the position of normal galaxies by shifting the UV-selected galaxies
by $\sim$ 0.15 dex in the vertical direction,
one can find deviations of oxygen abundance of intermediate-redshift
(z $\sim$ 0.4) galaxies as large as $\sim$ 0.1 dex (see Fig.~\ref{figure:zdoh})
if the correlation is real.
This could suggest that intermediate-redshift galaxies are
slightly less advanced in their evolution and, as a consequence, are more
metal-deficient (by $\sim$ 0.1 dex) than local galaxies. However, since
the dispersion of points in Fig.~\ref{figure:zdoh} is large (the mean value of
deviation $\Delta$log(O/H) $\sim$ 0.15dex) and the correlation is very weak, no 
firm conclusion can be drawn from this study on the evolution with redshift of 
the oxygen abundance of galaxies.

\section{Conclusions}

The validity of oxygen and nitrogen abundances derived via the $P$--method 
from the global spectra of galaxies has been investigated using a
collection of published spectra of individual H\,{\sc ii} regions in irregular
and spiral galaxies. It has been shown that the oxygen abundance derived
from the global emission-line spectrum of high-metallicity galaxies using
the $P$--method agrees very well with the oxygen abundance at galactocentric
distance ($r=0.4R_{\rm 25}$), traced by individual H\,{\sc ii} regions, if most
of H\,{\sc ii} regions belong to the upper branch of the O/H -- $R_{\rm 23}$
relation. The oxygen abundances in low-metallicity galaxies derived from the
global emission-line spectra via the $P$--method are slightly
underestimated, by $\Delta$(O/H) $\le$ 0.1 dex. We thus confirm the conclusions
of Kobulnicky, Kennicutt \& Pizagno (1999) that global emission-line spectra
can reliably indicate the chemical properties of galaxies.

It has been shown that the comparison of the global spectrum of a galaxy with a
collection of spectra of individual H\,{\sc ii} regions can be used to
distinguish high-- versus low--metallicity objects, i.e. to establish whether
the galaxy belongs to the upper or lower branch of the O/H -- $R_{\rm 23}$
relation. This method (called the $D$--method) can further be used for the determination of
accurate abundances in a galaxy.

The oxygen and nitrogen abundances in a sample of UV-selected local and
intermediate-redshift galaxies from Contini et al. (2002) have been determined
using both the $D$--method and the $P$--method. It has been found that
the UV-selected galaxies fill more or less uniformly the area outlined in the
N/O -- O/H diagram by individual H\,{\sc ii} regions of well-studied local
 galaxies. The UV-selected galaxies do not show any significant  shift
in the N/O -- O/H diagram relative to the individual H\,{\sc ii} regions
in ``normal'' local galaxies.

Finaly, it has been found that the intermediate-redshift galaxies seem to 
systematically deviate from the metallicity -- luminosity trend of local galaxies.
One explanation could be that intermediate-redshift galaxies are slightly
less advanced in their evolution and, as a consequence, are slightly more
metal-deficient than local galaxies of the same luminosity. As an alternative, 
intermediate-redshift galaxies may have just undergone a powerful starburst which
temporary increases their blue luminosity. Given the limited size of the present 
samples of intermediate-redshift galaxies, no firm conclusion can be drawn from 
this study on the evolution with redshift of the oxygen abundance of galaxies. 
No doubt, on-going deep spectroscopic surveys, such as VVDS or DEEP, 
will shed new light on this important issue.

\begin{acknowledgements}
We are grateful to the anonymous referee for making 
constructive suggestions to improve the paper.
L.S.P thanks the Laboratoire d'Astrophysique de l'Observatoire
Midi--Pyr\'{e}n\'{e}es (UMR 5572) for their hospitality during his visit where
this work was done. This study was supported (LSP) by the Laboratoire
d'Astrophysique de l'Observatoire Midi--Pyr\'{e}n\'{e}es (UMR 5572) within the
framework of ``postes rouges'' and by the Ukrainian Fund of Fundamental Investigation,
grant No 02.07/00132. JVM was supported by project AYA2001-3939-03-C01 of the 
spanish Programa Nacional de Astronomia y Astrofisica of the MCyT
\end{acknowledgements}

\end{document}